\documentclass[12pt,preprint]{aastex}
\usepackage{spr-astr-addons}
\usepackage{graphicx}

\newcommand{\HII}{{\ion{H}{2}}\,}

\newcommand{\OIIIHb}{[{\ion{O}{3}}]/H$\beta$}
\def\ratioR23{([\ion{O}{2}]~$\lambda$3727 +[\ion{O}{3}]~$\lambda\lambda$4959,5007)/H$\beta$}
\def\R23{${\rm R}_{23}$}

\newcommand{\NII}{[{\ion{N}{2}}]}

\newcommand{\NIIHa}{[\ion{N}{2}]/H$\alpha$}
\newcommand{\SIIHa}{[\ion{S}{2}]/H$\alpha$}

\newcommand{\SII}{[{\ion{S}{2}}]}

\def\O4363{[{\ion{O}{3}}]~$\lambda$4363}
\newcommand{\OIII}{[{\ion{O}{3}}]}

\newcommand{\Ha}{{H$\alpha$}}

\def\L60{L$_{60}$}

\begin{document}

\title{Chemical abundances in high-redshift galaxies: \\A powerful new emission line diagnostic}
\shorttitle{Star Formation in ULIRGs}
\shortauthors{Dopita et al.}

\author{ Michael A. Dopita\altaffilmark{1}\altaffilmark{2},  Lisa J. Kewley\altaffilmark{1}\altaffilmark{3},\newline Ralph S. Sutherland\altaffilmark{3}, \& David C. Nicholls\altaffilmark{1}}
\email{Michael.Dopita@anu.edu.au}
\altaffiltext{1}{Research School of Astronomy and Astrophysics, Australian National University, Cotter Rd., Weston ACT 2611, Australia }
\altaffiltext{2}{Astronomy Department, King Abdulaziz University, P.O. Box 80203, Jeddah, Saudi Arabia}
\altaffiltext{3}{Institute for Astronomy, University of Hawaii, 2680 Woodlawn Drive, Honolulu, HI 96822, USA}

\email{Michael.Dopita@anu.edu.au}
\begin{abstract}
This \emph{Letter} presents a new, remarkably simple diagnostic specifically designed to derive chemical abundances for high redshift galaxies. It uses only the H$\alpha$, \NII and \SII\ emission lines,  which can usually be observed in a single gating stetting, and is almost linear up to an abundance of $12+\log {\rm (O/H)} = 9.05$ . It can be used over the full abundance range encountered in high redshift galaxies. By its use of emission lines located close together in wavelength, it is also independent of reddening. Our diagnostic depends critically on the calibration of the N/O ratio. However, by using realistic stellar atmospheres combined with the N/O vs. O/H abundance calibration derived locally from stars and \HII\ regions, and allowing for the fact that high-redshift \HII\ regions have both high ionisation parameters \emph{and} high gas pressures, we find that the observations of high-redshift galaxies can be simply explained by the models without having to invoke arbitrary changes in N/O ratio, or the presence of unusual quantities of Wolf-Rayet stars in these galaxies.
\end{abstract}

\section{Introduction}
The chemical history of the universe provides a fossil record of the generations of star formation in galaxies, modulated by both inflow of pristine gas and by galactic-scale gas outflows.   The most commonly used method to probe the chemical history of the universe is to compare metallicity-sensitive optical emission-lines for ensembles of galaxies at different redshifts.  These investigations have led to new insights into the relationship between galaxy mass and metallicity at high-z \citep[][and references therein]{Zahid13,Zahid14,Izotov15} and new insights into the relationship between galaxy mass, metallicity, and star formation rate at high-z \citep[][and references therein]{Wuyts14,Sanders15,Maier15,Salim15,delosReyes15}. 

Regrettably, such abundance studies have been plagued by both observational and measurement uncertainties.   The commonly used metallicity calibrations may disagree by more than an order of magnitude \citep{Kewley06}.  This discrepancy has many potential causes, including calibration errors resulting from the use of local \HII\ regions relying on direct measurements of electron temperature $T_e$  \citep{Pilyugin05,Pilyugin11}, the use of hybrid approaches \citep{vanZee98}, relying only on strong emission line methods \citep{McGaugh91}, or simply by directly calibrating the theoretical photoionization models \citep{LopezSanchez12}. 

Current high-z metallicity calibrations assume that the ISM conditions, such as the ISM pressure and ionization parameter, are similar to those found in local galaxies.   These diagnostics are either based on samples of local \HII\ regions \citep{Denicolo02,Pettini04,PM09}, local galaxies \citep{Marino13,Morales-Luis14}, or photoionization models based on local galaxy conditions \citep{Kewley02,Blanc15}.

However, it is now clear that the conditions in the ISM of galaxies evolve with redshift.  Galaxies at high redshift have substantially larger \OIIIHb\ line ratios at high redshift ($z\sim 3$) than the present day \citep[e.g.,][and references therein]  {Hainline09,Bian10,Yabe12,Kewley13b,Holden14,Steidel14}. It is likely that a change in both ionization parameter and in the ISM pressure is responsible for this difference \citep{Brinchmann08,Liu08,Hainline09,Bian10,Nakajima12,Kewley13a}.  
Recently, \citet{Kewley15} showed that the ionization parameter changes systematically in galaxies between $0<z<0.6$, causing a rise in the \OIIIHb\ ratio, and a fall in the \NIIHa\ ratio.   Therefore, is crucial to use a metallicity calibration that either takes these parameters into account, or ideally is independent of both the ionization parameter and the pressure in the ISM. 

Recently \citet{Masters14} used the Magellan FIRE instrument to observe a sample of $z\sim2$ galaxies in which all of the important diagnostic lines are observed, and \citet{Shapley15} has extended this sample using galaxies drawn from the MOSFIRE Deep Evolution Field (MOSDEF) Survey. Both of these authors find that their data points are offset in the BPT diagram toward higher \OIIIHb\ for a given \NIIHa, as had already been noted in $z\sim1$ star-forming galaxies. However, composite spectra derived from the samples do not show a corresponding offset from the local star-forming sequence on the \OIIIHb\ vs. \SIIHa\ diagram. Both of these authors interpret this result in terms of an enhanced population of Wolf-Rayet stars at high redshift, leading to an enhancement in the relative N abundance, and a much hotter ionising spectrum (modelled by \citet{Shapley15} as a hot Black Body). We will critically examine this hypothesis in this Letter, and show that it has a simpler explanation.

Here, we present a new metallicity calibration which relies solely on the red  \Ha, \NII, and \SII\ emission-lines.  This metallicity calibration is effectively independent of both ionization parameter and ISM pressure, and is valid over the full range of abundance encountered in high redshift galaxies.  The line ratios used do not require either flux calibration, or extinction correction.  We anticipate that this calibration will be critical for high-z metallicity studies where different lines usually have to be observed in different bands, and where the ionization parameter and/or the ISM pressure may be significantly different from those in local galaxies.

\section{Models}
We have used the {\tt Mappings 5.0} code (Sutherland et al. 2015, in prep.) {\footnote{Available at {\tt miocene.anu.edu.au/Mappings}} to construct a grid of photoionisation models for \HII\ regions. This code is the latest version of the {\tt Mappings 4.0} code described in \citep{Dopita13}, and includes numerous upgrades to both the input atomic physics and the methods of solution.

The spectrum of an \HII\ region depends upon a number of fundamental parameters. Of course the chemical abundance set is the most important amongst these, and since the cooling of the \HII region is moderated by the gas-phase abundances, it is important to have a reasonable model for the dust grain content of the \HII regions.  The dust grains are also very important in determining the photoelectric heating of the plasma \citep{Dopita00}. The excitation of the nebula was generally thought to be moderated by three parameters, the ionisation parameter $\log U$, the shape of the cluster EUV spectrum, and the pressure in the ionised plasma. However, \citet{Dopita14} showed that the pressure in the ionised gas is also impotent. Here, for the first time, we systematically investigate the effect of this pressure parameter on the strong-line spectra of \HII\ regions, and demonstrate that this parameter provides a ``missing link'' which finally allows us to explain the peculiarities of the spectra of high-redshift galaxies.

Recently, \citet{Nicholls12} have suggested that the electrons in \HII\ regions are characterised by a $\kappa-$distribution in energy, rather than by a simple Maxwell-Boltzmann (M-B) distribution. The effect of this on the emission line spectra of \HII\ regions was systematically investigated by \citet{Dopita14}. Here we present grids with the M-B distribution only, but we have also computed the effects of a $\kappa-$distribution. As we show in Section \ref{Sect4}, this makes almost no difference to our new abundance calibration presented here.

In these models we have adopted the local galactic concordance (LGC) abundances based upon the \citet{Nieva12} data on early B-star data. These have the advantage that they sample the abundances in the local region of the galaxy (out to 500 pc), providing the current abundances in this region. \citet{Nieva12} gave the N/O and C/O ratios in the range $5.9 < 12+ \log {\rm{O/H} < 9.0}$. The determination of these ratios is vital to photoionisation models, since N and C are important coolants in the nebula, are both in part secondary nucleosynthesis elements, and are consequently difficult to directly calibrate. In particular, C is only readily observable in the UV, and may be locked up in dust (which cannot exist in B-star atmospheres).  The  \citet{Nieva12} data provides the abundances of the main coolants, H, He, C, N. O, Ne, Mg, Si and Fe. For the light elements we use the \citet{Lodders09} abundance, while for all other elements the abundances are based upon \citet{Scott15a, Scott15b} and \citet{Grevesse10}. Thus, in the LGC scale, the ``local region'' reference abundance has $12+\log {\rm (O/H)} = 8.77$, as opposed to the \citet{Grevesse10} solar value of $12+\log {\rm (O/H)} = 8.69$. 

{ Since the calibration presented in this \emph{Letter} depends critically on the assumed calibration of N/O vs. O/H, we have presented the calibration we used in Figure \ref{fig1}, takeen from \citet{Nicholls15}. This calibration is based on a mixture of both stellar and nebular sources \citep{Izotov99, Israelian04, Spite05, Nieva12}. The global enrichment pattern can be represented by mixture of primary and secondary nucleosythesis, as can be seen clearly in Figure \ref{fig1}. This point will be discussed in more detail in Section \ref{N/O}.}

\begin{figure}[ht]
\includegraphics[width=1.0\hsize]{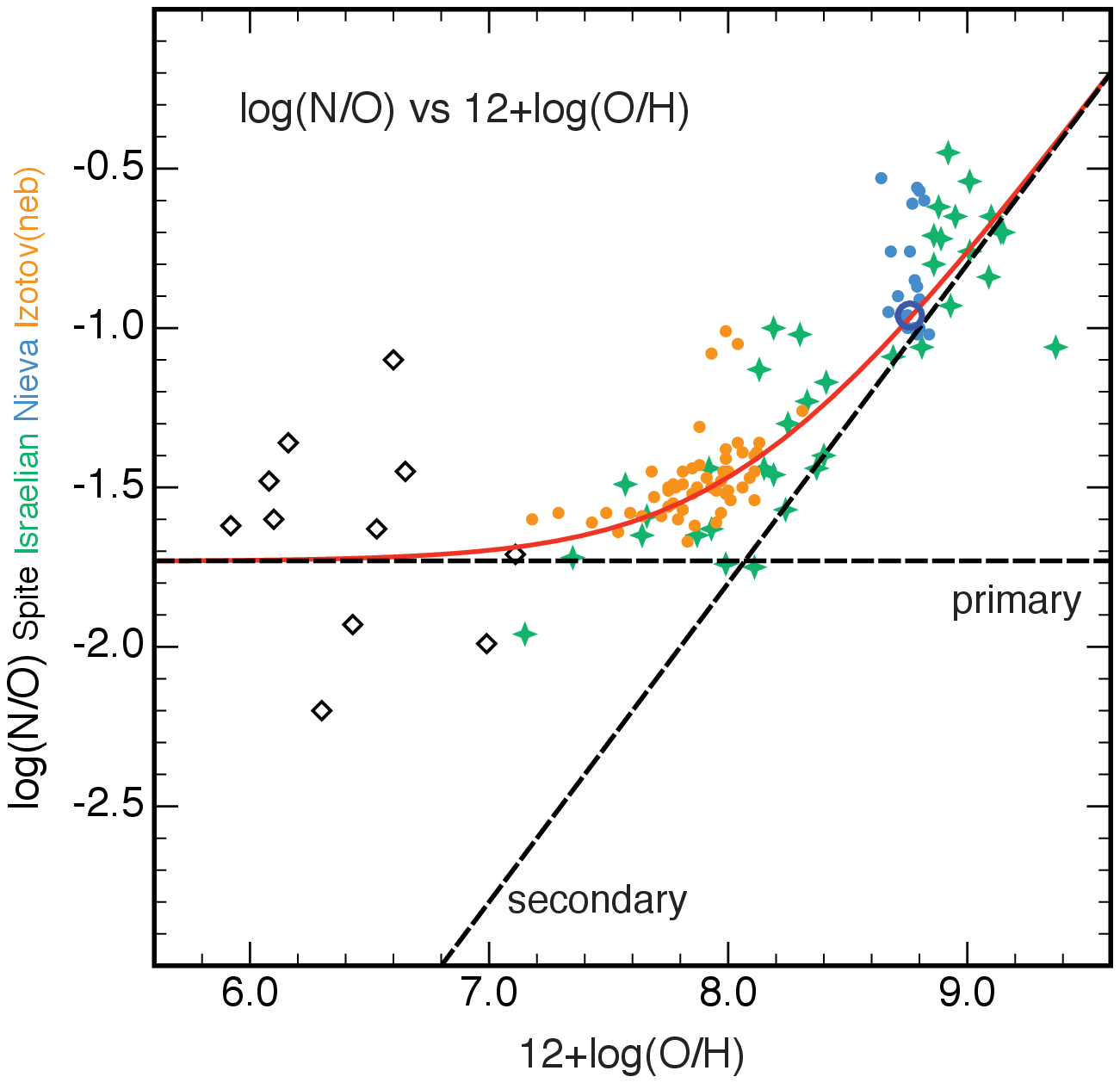}
\caption{The calibration of N/O vs. O/H used in this paper. The data points are derived from the following papers: \citet{Izotov99, Israelian04, Spite05} and \citet{Nieva12}}\label{fig1}
\end{figure}

The depletion factors of the heavy elements onto dust are derived from the formulae of \citet{Jenkins09}, extended to the other elements on the basis of condensation temperatures and/or position on the periodic table. We have investigated the effect of changing the Fe depletion,  $\log D_{\rm Fe}$ (in the range $-1.0 > \log D_{\rm Fe} > -2.5$, and have ascertained that it makes no difference to abundance diagnostics at solar abundance, but may introduce an uncertainty of $\pm 0.12$\,dex. at the extremities of the abundance range computed here.

For the EUV cluster spectra we use the {\tt Starburst99} models \citep{SB99} interpolated in abundance as described in \citep{Dopita13}, and for these we have computed grids of spherical, isobaric \HII regions at 3.0, 2.0, 1.0, 0.5, 0.3, 0.2 and 0,1 times this ``local region'' reference abundance, and covering an ionisation parameter at the inner edge of the nebula of $-3.5 <\log U <-2.0$ in steps of 0.25. Because the pressure in the \HII\ region is an important parameter which changes the excitation in the nebula in a similar way the the ionisation parameter \citep{Dopita14}, we have computed grids at four pressures, $\log P/k = 5.2, 5.7, 6.2$ and 6.7\,cm$^{-3}$K, which cover the full range commonly encountered in the \HII\ regions of normal galaxies and starburst galaxies in the local universe \citep{Dopita14}. At high redshifts,  high values of pressure also seem to be appropriate. For example, \citet{Masters14} reports electron densities in their $z\sim2$ sample of $100 \lesssim n_e \lesssim 400$, which corresponds to $6.2 \lesssim \log P/k \lesssim 6.7$\,cm$^{-3}$K.

\section{Results}
It is very difficult to obtain meaningful abundances using a single line ratio of a forbidden line and a Hydrogen recombination line, since all such ratios are two-valued in abundance space, and are grossly affected by both ionisation parameter and pressure. Even with two emission line ratios it is difficult, since the grids of models tend to fold up on themselves \citep{VO87,Dopita13}. Generally speaking, the strategy here has been to search for a line ratio which depends principally on abundance, and a second which is sensitive to the excitation, for example \citet{McGaugh91,Kewley02,Kobulnicky04}. Frequently, however the lines used are far apart in wavelength, reddening corrections large, and all lines required may not be available to be observed in the case of high-redshift galaxies. 

\citet{Vogt14} demonstrated the utility of classification of galaxies using 3D emission line ratio diagrams, and this is the approach we have adopted here. Specifically, we have searched amongst the set of commonly-used strong line ratios, choosing 3 sets of line ratios which are either known to be excitation dependent, or else more abundance-sensitive. We also applied the criterion that the emission line ratios used be close together in wavelength to eliminate, as far as possible, the effects of differential reddening. We then rotated the theoretical grids in pitch, yaw and roll in order to separate the abundance sensitivity into one axis of projection, and the excitation sensitivity in the other axis. 

We discovered that the use of the \NII $\lambda 6484$/H$\alpha$ the  \NII $\lambda 6484$/\SII $\lambda \lambda 6717,31$ and the \OIII $\lambda 5007$/H$\beta$ line ratios provided an excellent result. The first two ratios are abundance sensitive, and the third is more sensitive to excitation. The result is shown in Figure \ref{fig2}, where the $y-$axis gives the abundance (with very small residual sensitivity to either $\log U$ or $\log P/k$, while the $x-$axis demonstrates sensitivity to both ionisation parameter and pressure.

{ In Figure \ref{fig2}, apart from the high-redshift (mean) points from the \citet{Masters14} and \citet{Shapley15} samples, we have used the observations of individual \HII\ regions from \citet{vanZee98}. We have chosen to use the van Zee sample rather than the SDSS nuclear data to avoid the aperture effects, the inclusion of diffuse emission, and the bias towards high abundance inherent in the SDSS sample. In addition, the van Zee sample represents a homogenenous sample of integral \HII\ region spectra, observed and reduced in the same way, with the same instrument. This minimises systematic errors. In addition, it provides conditions in the ionised gas which are more comparable to the high-redshift galaxies because:
\begin{enumerate}
\item{It samples HII regions across the faces of galaxies, ensuring that a wide range 
of metallicities and ionisation parameters are sampled. In particular, HII regions 
with abundances similar to the high-redshift sample are well-represented.}
\item{By its choice of high surface brightness and high luminosity HII regions, the van Zee
sample is biased towards selecting HII regions with both high local specific star formation rates,
and with high ISM pressure, both of which are believed to apply to high-redshift galaxies.}
\end{enumerate}
}

\begin{figure}[ht]
\includegraphics[width=1.0\hsize]{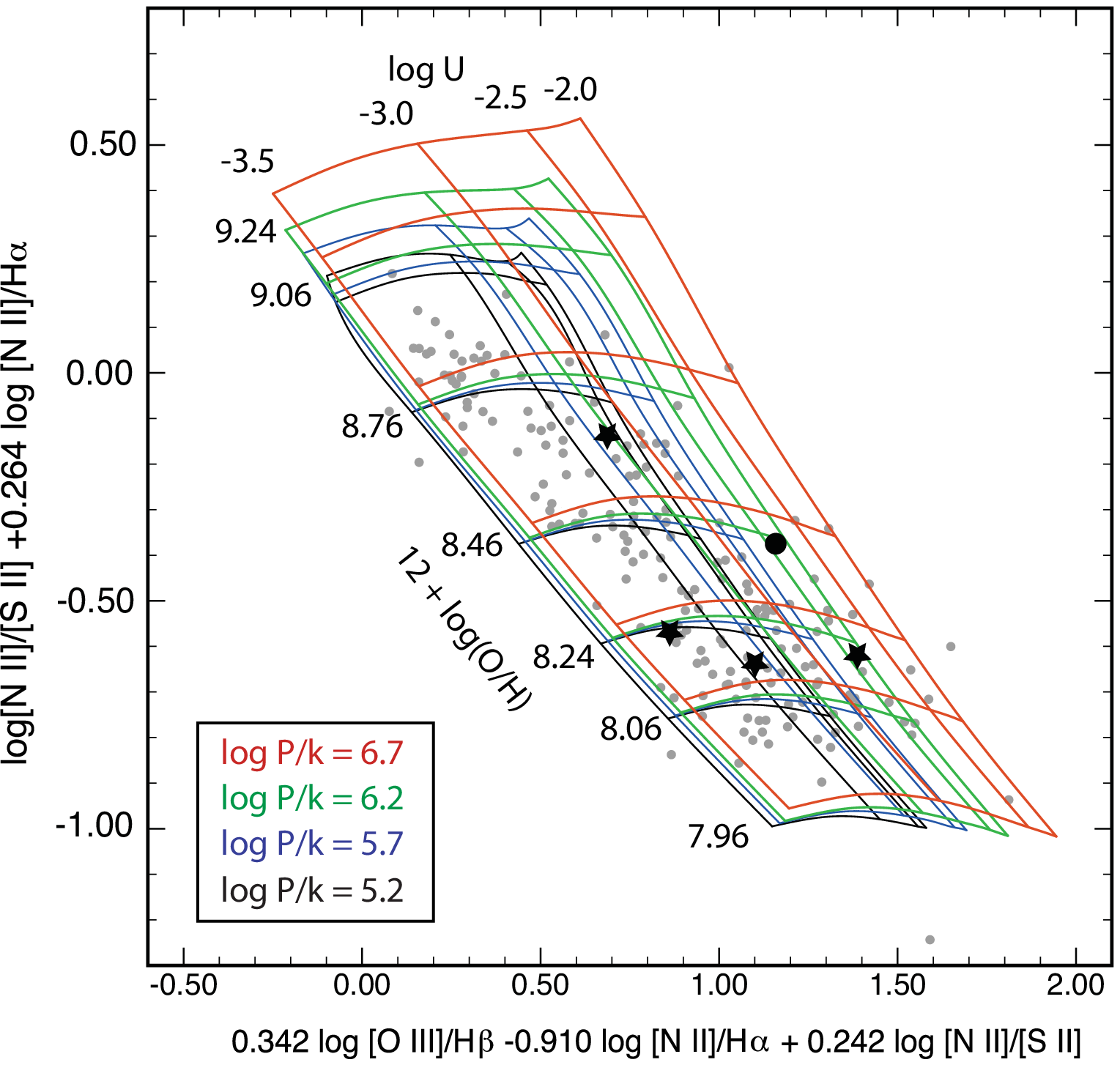}
\caption{The 3D projection of the  \NII $\lambda 6484$/H$\alpha$ the  \NII $\lambda 6484$/\SII $\lambda \lambda 6717,31$ and the \OIII $\lambda 5007$/H$\beta$ line ratios which cleanly separates the abundance ($y-$axis) from the effects of pressure and ionisation pressure ($x-$axis). For comparison with the theoretical grid, the observations of \citet{vanZee98} of individual \HII regions in local spiral galaxies are shown as grey circles, and for high-redshift galaxies we plot the $z\sim2$ MOSDEF stacks from \citet{Shapley15} (stars), and the $z\sim2$ composite spectrum from the Magellan FIRE survey from \citet{Masters14} (filled black circle). These data points are generally consistent with sub-solar metallicity, high pressure, and high ionisation parameter.}\label{fig2}
\end{figure}

 From Figure \ref{fig2} it is evident that \emph{only} the line ratios \NII $\lambda 6484$/H$\alpha$ the  \NII $\lambda 6484$/\SII $\lambda \lambda 6717,31$ need to be used to obtain a good estimate of the chemical abundance. Our use of only the red lines allows observers to effectively ignore reddening corrections. Indeed, these lines may be the only ones available to be observed at certain red-shifts. 
 
 This emission line combination has notable advantages over the calibration of \citet{Pettini04}, which uses the \OIII $\lambda 5007$/H$\beta$ and  \NII $\lambda 6484$/H$\alpha$ line ratios, through the compound so-called O3N2 ratio introduced by \citet{Alloin79}. For high-redshift galaxies, the use of this requires observations in two wavelength bands, and is strongly affected by both the pressure and ionisation parameter issues discussed above. 
 
 \begin{figure}[ht]
\includegraphics[width=0.9\hsize]{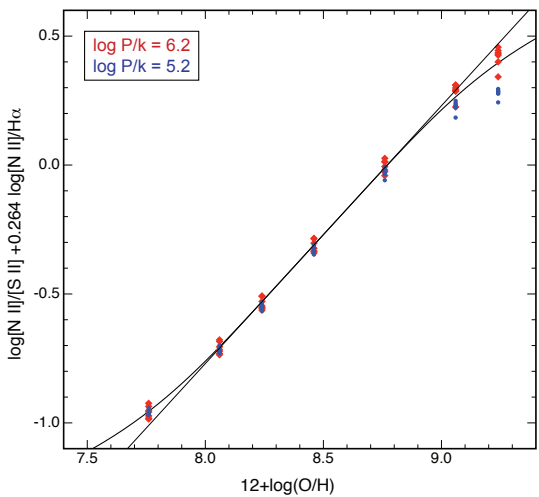}
\caption{Calibration of the  \NII/\SII  and  \NII/H$\alpha$ emission line ratio combination against $12+\log {\rm (O/H)}$. The red group of points represent the models having $\log P/k = 6.2$\,cm$^{-3}$K, while the blue points are for $\log P/k = 5.2$\,cm$^{-3}$K. The lines shown are the best linear fits and 5th. order fits as described in the text.}\label{fig3}
\end{figure}

 For these two line ratios the calibration to abundance is particularly simple. This is shown graphically in Figure \ref{fig3}. A linear fit is good up to $12+\log {\rm (O/H)} \sim 9.05$ (within the limits imposed by the uncertainties in the models themselves, including the uncertainty in the depletion factor discussed above). 
With:  
\begin{eqnarray}
y  =    \log {\rm [N II]/[S II]} + 0.264 \log {\rm  [N II] / H}\alpha, \\  
12 + \log{\rm (O/H)}  =  8.77 +y
\end{eqnarray}
This linear fit is shown on Figure \ref{fig2}. If so desired, a somewhat improved fit can be got by adding a 5th. order correction term:
\begin{equation}
12 + \log{\rm (O/H)}  =  8.77 +y +0.45(y+0.3)^5,
\end{equation}
which is shown as the curved line on Figure \ref{fig3}. However, other errors both observational and theoretical are likely to mask the effect of such a correction.

\section{Discussion}\label{Sect4}
\subsection{The new diagnostic}
From an investigation of triplets of emission line ratios in 3D, we have discovered a simple, linear diagnostic with a wide dynamic range which uses only the red lines; H$\alpha$,  \NII $\lambda 6484$ and the \SII\ doublet at $\lambda \lambda 6717,31$ to determine the O/H ratio. This should prove very valuable in the investigation of the chemical evolution of the Universe, as frequently these lines are the only ones observable in high redshift galaxies, all can be observed together with a single instrument setting, and reddening corrections are sufficiently small to be neglected.

\subsection{Calibration of the N/O ratio}\label{N/O}
Because both of the ratios used are dependent on the Nitrogen abundance, clearly our calibration is strongly dependent on the correct calibration of the N/O ratio with O/H. In particular, Nitrogen behaves in part as a secondary nucleosynthetic element because at low abundance N is promptly enriched as a result of CN processing and mass-loss in massive stars, while at later epochs it is more produced by hot-bottom burning in intermediate-mass stars. For low-abundance stars (appropriate to the high-redshift regime) the effect of hot-bottom burning has recently been investigated by \citet{Fishlock14}. Recently \citet{Nicholls15} has examined the calibration of the N/O vs. O/H calibration using both \HII\ regions and results from old stars in the Galaxy (galactic archeology). These authors used data from \cite{Spite05}derived from halo metal-poor unmixed giants, \citet{Fabbian09} from halo solar type dwarfs and sub-giants, and from \citet{Nieva12} for the local B stars. In addition, they used data derived from Blue Compact Galaxies by \citet{Izotov99}. These authors point out that there is little evidence for dust in these objects, and, by implication, that there is relatively little oxygen or nitrogen depletion into dust. Together, these data are remarkably consistent, and provide the tight calibrate which we have used in this paper. This calibration is in fair agreement with that of \citet{PM13} for intermediate-redshift galaxies which, however, display a broader scatter in N/O at the low-abundance end.

Sulphur, like Oxygen is an $\alpha-$process element, and both are enriched in a similar way and on a similar timeframe. Because the time delay required to transition from primary to secondary Nitrogen production, the N/O ratio may be affected by scatter at a given O/H ratio, depending on the rate of star formation enrichment rate, and the mass of the parent galaxy \citep{Pettini02, PM09, PM13}.  

The recent observations by \citet{Masters14} and \citet{Shapley15} have cast some doubt on the validity of the N/O vs. O/H calibration (although regrettably they did not give the actual scaling they used in their models). These authors find that  \OIIIHb\  is elevated above what can simply be explained in terms of an elevated ionisation parameter. Furthermore, on the Phillips Baldwin and Terlevich (BPT) diagrams they find that \NIIHa\ is offset towards higher values relative to \SIIHa . They tentatively identify this as the effect of hot Wolf-Rayet stars with high effective temperatures ($T_e \sim 80000$ \,K), required to produce high \OIIIHb\ ) and which enrich the surrounding ISM with N (producing the high observed \NIIHa).

Is this hypothesis correct?  A major problem with it is that the important effects of pressure on the \HII\ region spectrum is not taken into account. This has a strong influence on the emission line ratios used in the BPT diagrams. This effect was initially discussed in the context of starburst galaxies by \citet{Dopita14}, who showed that pressure had a similar effect to ionisation parameter, at least at moderate pressures. The models presented here are the first which investigate the effect of interstellar pressure up to the range observed in these high-redshift galaxies ($100 \lesssim n_e \lesssim 400$, or $6.2 \lesssim \log P/k \lesssim 6.7$\,cm$^{-3}$K. This has two principal effects. First, the \OIIIHb\ ratio becomes elevated. This is due to the suppression of fine-structure cooling in the far-IR due to higher electron density, leading to higher nebular temperature, and stronger \OIII\ emission. Second, the \SIIHa\ ratio is decreased. This is the result of collisional de-excitation of the \SII\ doublet. 

These effects are clearly shown in Figure \ref{fig4}. Here the displacement remarked upon by \citet{Masters14} and \citet{Shapley15}between the local \HII\ regions and the high redshift composite spectra is clear on the \NIIHa\ vs. \OIIIHb\ diagnostic, while the \SIIHa\ vs. \OIIIHb\ diagnostic show no such offset. On the theoretical grids, it is clear that this can be explained by the ``wrap round'' in the grids, which is very marked in the case of the  \SIIHa\ vs. \OIIIHb\ diagnostic. However, on both grids, the high-redshift points are consistent with sub-solar metallicity, high ionisation parameter, and also the high gas pressures implied by the observed \SII\ $\lambda\lambda6717/6731$line ratios \citep{Masters14}. Precisely the same conclusions are obtained by inspection of Figure \ref{fig2}. 

Therefore, using realistic stellar atmospheres combined with the observed N/O vs. O/H abundance calibration, and allowing for the fact that high-redshift \HII\ regions have both high ionisation parameters \emph{and} high gas pressures, we find that the observations of high-redshift \HII\ regions can be explained without having to invoke arbitrary changes in N/O ratio, or the presence of unusual quantities of Wolf-Rayet stars.

 \begin{figure*}[ht]
\includegraphics[width=1.0\hsize]{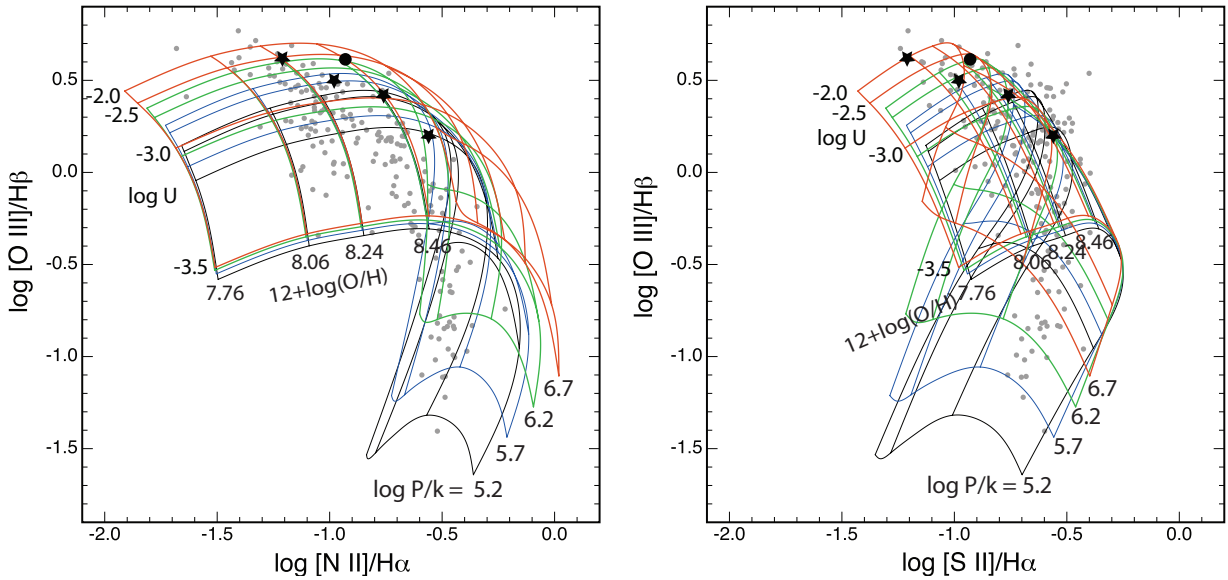}
\caption{BPT diagrams showing the effect of pressure on the shape of the theoretical grids. The meaning of the symbols is the same as in Figure \ref{fig2}. For clarity, the values of $\log U$ are only marked for the case of the high-pressure grid. Note that a number of the \citet{vanZee98} \HII\ regions lie in the high pressure, high ionisation parameter regime, particularly those of lower abundance. Note also that the high-redshift galaxies are also consistent with having high pressure, high ionisation parameters and sub-solar metallicities in the \HII\ region  - in agreement with Figure \ref{fig2}. Clearly, there is no need to increase the relative N abundance in order to explain the high-redshift galaxy observations by \citet{Masters14} and \citet{Shapley15}.}\label{fig4}
\end{figure*}

Other issues which may affect our calibration are that the EUV spectra we have used may be inaccurate at low metallicities, or that the electrons in the nebula have a $\kappa-$distribution, rather than a simple Maxwell-Boltzmann distribution \citep{Nicholls12, Dopita14}. We have estimated the effect in the case of a $\kappa-$distribution, by running a grid models with $\kappa = 20$ and  $\log P/k = 5.7$\,cm$^{-3}$K.  We determined that in this case, the zero point in equation 1 is raised from 8.77 to 8.79 -- a remarkably small difference. The slope of the relationship is increased by about 3\%, and becomes almost exactly linear with abundance.

With these caveats, we are confident that our new abundance diagnostic will prove of great value to those who are attempting to probe the chemical evolution of the high-redshift Universe.

\begin{acknowledgments}
Dopita and Kewley acknowledge the support of the Australian Research Council (ARC) through Discovery project DP130103925. Dopita would also like to thank the Deanship of Scientific Research (DSR), King AbdulAziz University for additional financial support as Distinguished Visiting Professor under the KAU Hi-Ci program.
\end{acknowledgments}


\begin{thebibliography}{}
\bibitem[Alloin et al.(1979)]{Alloin79}Alloin D., Collin-Souffrin S.,  Joly M., \&  Vigroux L., 1979, \aap, 78, 200

\bibitem[{{Bian} {et~al.}(2010){Bian}, {Fan}, {Bechtold}, {McGreer}, {Just},  {Sand}, {Green}, {Thompson}, {Peng}, {Seifert}, {Ageorges}, {Juette},  {Knierim}, \& {Buschkamp}}]{Bian10} {Bian}, F. {et~al.} 2010, \apj, 725, 1877

\bibitem[{{Blanc} {et~al.}(2015){Blanc}, {Kewley}, {Vogt}, \& {Dopita}}]{Blanc15}{Blanc}, G.~A., {Kewley}, L., {Vogt}, F.~P.~A., \& {Dopita}, M.~A. 2015, \apj, 798, 99

\bibitem[{{Brinchmann} {et~al.}(2008){Brinchmann}, {Pettini}, \&  {Charlot}}]{Brinchmann08} {Brinchmann}, J., {Pettini}, M., \& {Charlot}, S. 2008, \mnras, 385, 769.

\bibitem[{{de los Reyes} {et~al.}(2015){de los Reyes}, {Ly}, {Lee}, {Salim},  {Peeples}, {Momcheva}, {Feddersen}, {Dale}, {Ouchi}, {Ono}, \& {Finn}}]{delosReyes15}{de los Reyes}, M.~A. {et~al.} 2015, \aj, 149, 79

\bibitem[{{Denicol{\' o}} {et~al.}(2002){Denicol{\' o}}, {Terlevich}, \&  {Terlevich}}]{Denicolo02}{Denicol{\' o}}, G., {Terlevich}, R., \& {Terlevich}, E. 2002, \mnras, 330, 69

\bibitem[Dopita \& Sutherland(2000)]{Dopita00}Dopita, M.~A., \& Sutherland, R.~S. 2000, \apj, 539, 742.

\bibitem[Dopita et al.(2013)]{Dopita13} Dopita, M.~A., Sutherland, R.~S., Nicholls, D.~C., Kewley, L.~J., \& Vogt, F.~P~A. 2013, \apjs,  208, 10

\bibitem[Dopita et al.(2014)]{Dopita14} Dopita, M.~A., Rich, J., Vogt, F.~P.~A.; Kewley, L.~J., Ho, I.-T,  Basurah, H.~M., Ali, A. \& Amer, M.~A. 2014, \apss, 350, 741

\bibitem[Fabbian et al.(2009)]{Fabbian09}Fabbian D., Nissen P.~E., Asplund M., Pettini M., \& Akerman C., 2009, \aap, 500, 1143

\bibitem[Fishlock et al.(2014)]{Fishlock14}Fishlock, C.~K., Karakas, A.~I., Lugaro, M. \& Yong, D. 2014, \apj, 797, 44

\bibitem[Grevesse et al.(2010)]{Grevesse10} Grevesse, N., Asplund, M., Sauval, A. J., \& Scott, P. 2010, \apss, 328, 179

\bibitem[{{Hainline} {et~al.}(2009){Hainline}, {Shapley}, {Kornei}, {Pettini},  {Buckley-Geer}, {Allam}, \& {Tucker}}]{Hainline09}{Hainline}, K.~N., {Shapley}, A.~E., {Kornei}, K.~A., {Pettini}, M., {Buckley-Geer}, E., {Allam}, S.~S., \& {Tucker}, D.~L. 2009, \apj, 701, 52

\bibitem[{{Holden} {et~al.}(2014){Holden}, {Oesch}, {Gonzalez}, {Illingworth},  {Labbe}, {Bouwens}, {Franx}, {van Dokkum}, \& {Spitler}}]{Holden14} {Holden}, B.~P. {et~al.} 2014, ArXiv e-prints, 1401.5490

\bibitem[Israelian et al.(2004)]{Israelian04} Israelian, G., Ecuvillon, A., Rebolo, R., Garc'a-L\'opez, R., Bonifacio, P., \& Molaro, P., 2004, \aap, 421, 649

\bibitem[Izotov et al.(1999)]{Izotov99}Izotov Y.~I., \& Thuan T.~X., 1999, \apj, 511, 639

\bibitem[{{Izotov} {et~al.}(2015){Izotov}, {Guseva}, {Fricke}, \&  {Henkel}}]{Izotov15} {Izotov}, Y.~I., {Guseva}, N.~G., {Fricke}, K.~J., \& {Henkel}, C. 2015,  \mnras, 451, 2251.

\bibitem[Jenkins(2009)]{Jenkins09}Jenkins, E.~B. 2009, \apj, 700, 1299.

\bibitem[Kewley \& Dopita(2002)]{Kewley02}Kewley, L.~J. \& Dopita, M.~A. 2002, \apjs, 142, 35.

\bibitem[Kewley et al.(2006)]{Kewley06}Kewley, L. J., Groves, B., Kauffmann, G., \& Heckman, T. 2006, \mnras, 372, 961 

\bibitem[{{Kewley} \& {Ellison}(2008)}]{Kewley08} {Kewley}, L.~J., \& {Ellison}, S.~L. 2008, \apj, 681, 1183

\bibitem[{{Kewley} {et~al.}(2013{\natexlab{a}}){Kewley}, {Dopita}, {Leitherer},  {Dav{\'e}}, {Yuan}, {Allen}, {Groves}, \& {Sutherland}}]{Kewley13a}{Kewley}, L.~J., {Dopita}, M.~A., {Leitherer}, C., {Dav{\'e}}, R., {Yuan}, T.,  {Allen}, M., {Groves}, B., \& {Sutherland}, R. 2013{\natexlab{a}}, \apj, 774,  100.

\bibitem[{{Kewley} {et~al.}(2013{\natexlab{b}}){Kewley}, {Maier}, {Yabe},  {Ohta}, {Akiyama}, {Dopita}, \& {Yuan}}]{Kewley13b} {Kewley}, L.~J., {Maier}, C., {Yabe}, K., {Ohta}, K., {Akiyama}, M., {Dopita},  M.~A., \& {Yuan}, T. 2013{\natexlab{b}}, \apjl, 774, L10.

\bibitem[{{Kewley} {et~al.}(2015){Kewley}, {Zahid}, {Geller}, {Dopita}, \&  {Hwang}}]{Kewley15} {Kewley}, L.~J., {Zahid}, H.~J., {Geller}, M.~J., {Dopita}, M.~A., \& {Hwang},  H.~S. 2015, \apjl (in press)

\bibitem[Kobulnicky \& Kewley(2004)]{Kobulnicky04}Kobulnicky H.~A. \& Kewley L.~J. 2004, \apj, 617, 240

\bibitem[Leitherer et al.(1999)]{SB99}Leitherer, C., Schaerer, D., Goldader, J.~D., Delgado, R.~M.~G., et al. 1999, \apjs, 123, 3.

\bibitem[{{Liu} {et~al.}(2008){Liu}, {Shapley}, {Coil}, {Brinchmann}, \& {Ma}}]{Liu08} {Liu}, X., {Shapley}, A.~E., {Coil}, A.~L., {Brinchmann}, J., \& {Ma}, C.-P.  2008, \apj, 678, 758

\bibitem[{Lodders} \& {Palme}(2009)]{Lodders09} {Lodders}, K., \& {Palme}, H. 2009,  Meteoritics and Planetary Science Supplement, 72, 5154.
\bibitem[L\'opez-S\'anchez et al.(2012)]{LopezSanchez12}L\'opez-S\'anchez, \'A.~R.; Dopita, M.~A., Kewley, L.~J., Zahid, H.~J., Nicholls, D.~C., \& Scharw\"achter, J. 2012, \mnras, 426,2630

\bibitem[McGaugh(1991)]{McGaugh91}McGaugh, S.~S. 1991, \apj, 380, 140.

\bibitem[{{Maier} {et~al.}(2015){Maier}, {Ziegler}, {Lilly}, {Contini},  {P{\'e}rez-Montero}, {Lamareille}, {Bolzonella}, \& {Le Floc'h}}]{Maier15} {Maier}, C., {Ziegler}, B.~L., {Lilly}, S.~J., {Contini}, T.,  {P{\'e}rez-Montero}, E., {Lamareille}, F., {Bolzonella}, M., \& {Le Floc'h},
  E. 2015, \aap, 577, A14.

\bibitem[{{Marino} {et~al.}(2013){Marino}, {Rosales-Ortega}, {S{\'a}nchez},  {Gil de Paz}, {V{\'{\i}}lchez}, {Miralles-Caballero}, {Kehrig},  {P{\'e}rez-Montero}, {Stanishev}, {Iglesias-P{\'a}ramo}, {D{\'{\i}}az},   {Castillo-Morales}, {Kennicutt}, {L{\'o}pez-S{\'a}nchez}, {Galbany}, {Garc{\'{\i}}a-Benito}, {Mast}, {Mendez-Abreu}, {Monreal-Ibero}, {Husemann},  {Walcher}, {Garc{\'{\i}}a-Lorenzo}, {Masegosa}, {Del Olmo Orozco},  {Mour{\~a}o}, {Ziegler}, {Moll{\'a}}, {Papaderos},  {S{\'a}nchez-Bl{\'a}zquez}, {Gonz{\'a}lez Delgado}, {Falc{\'o}n-Barroso}, {Roth}, {van de Ven}, \& {Califa Team}}]{Marino13} {Marino}, R.~A. {et~al.} 2013, \aap, 559, A114.

\bibitem[Masters et al.(2014)]{Masters14}Masters, D., McCarthy, P., Siana, B., et al. 2014, \apj, 785, 1

\bibitem[{{Morales-Luis} {et~al.}(2014){Morales-Luis}, {P{\'e}rez-Montero},  {S{\'a}nchez Almeida}, \& {Mu{\~n}oz-Tu{\~n}{\'o}n}}]{Morales-Luis14} {Morales-Luis}, A.~B., {P{\'e}rez-Montero}, E., {S{\'a}nchez Almeida}, J., \&  {Mu{\~n}oz-Tu{\~n}{\'o}n}, C. 2014, \apj, 797, 81.

\bibitem[{{Nakajima} {et~al.}(2012){Nakajima}, {Ouchi}, {Shimasaku},  {Hashimoto}, {Ono}, \& {Lee}}]{Nakajima12} {Nakajima}, K., {Ouchi}, M., {Shimasaku}, K., {Hashimoto}, T., {Ono}, Y., \& {Lee}, J.~C. 2012, ArXiv e-prints, 1208.3260.

\bibitem[Nicholls et al.(2012)]{Nicholls12}Nicholls, D.~C., Dopita, M.~A. \& Sutherland, R.~S., 2012, \apj, 752, 148

\bibitem[Nicholls et al.(2015)]{Nicholls15}Nicholls, D.~C., Sutherland, R.~S., Dopita, M.~A. \& Kewley, L.~J. 2015, (in preparation).

\bibitem[{Nieva} \& {Przybilla}(2012)]{Nieva12}{Nieva}, M.-F., \& {Przybilla}, N. 2012, \aap, 539, A143

\bibitem[ P\'erez-Montero et al(2009)]{PM09}P\'erez-Montero, E., \& Contini, T 2009, \mnras, 398, 949

\bibitem[ P\'erez-Montero et al(2013)]{PM13}P\'erez-Montero, E., Contini, T, Lamareille F. , Maier, C. et al. 2013, \aap, 549, A25

\bibitem[Pettini et al(2002)]{Pettini02}Pettini M., Rix S.~A., Steidel C.~C., Adelberger K.~L., Hunt M.~P., \& Shapley A.~E., 2002b, \apj, 569, 742

\bibitem[Pettini \& Pagel(2004)]{Pettini04}Pettini, M. \& Pagel, B.~E.~J. 2004, MNRAS, 348, L59

\bibitem[Pilyugin \& Thuan(2005)]{Pilyugin05}Pilyugin, L.S., \& Thuan, T.~X.,  2005, \apj, 631, 231

\bibitem[Pilyugin \& Mattsson(2011)]{Pilyugin11}Pilyugin, L.S., \& Mattsson, L., 2011a, \mnras, 412, 1145

\bibitem[{{Salim} {et~al.}(2015){Salim}, {Lee}, {Dav{\'e}}, \&  {Dickinson}}]{Salim15} {Salim}, S., {Lee}, J.~C., {Dav{\'e}}, R., \& {Dickinson}, M. 2015, ArXiv  e-prints, 1506.03080.

\bibitem[{{Sanders} {et~al.}(2015){Sanders}, {Shapley}, {Kriek}, {Reddy},  {Freeman}, {Coil}, {Siana}, {Mobasher}, {Shivaei}, {Price}, \& {de  Groot}}]{Sanders15}{Sanders}, R.~L. {et~al.} 2015, \apj, 799, 138, 1408.2521.

\bibitem[Scott et al.(2015a)]{Scott15a}Scott, P., Asplund, M., Grevesse, N., Bergemann, M., \& Sauval, A. J. 2015a, \aap, 573, A26.

\bibitem[Scott et al.(2015b)]{Scott15b}Scott, P., et al. 2015b, \aap, 573, A25

\bibitem[Shapley et al.(2015)]{Shapley15}Shapley, A.~E., Reddy, N.~A., Kriek, M et al., 2015, \apj, 801

\bibitem[Spite et al.(2005)]{Spite05}Spite, M., Cayrel, R., Plez, B et al., 2005, \aap, 430, 655

\bibitem[van Zee et al.(1998)]{vanZee98}van Zee, L., Salzer, J.~J., Haynes, M.~P., O\'  ~Donoghue, A.~A., \& Balonek, T.~J., 1998, \aj, 116, 2805

\bibitem[{{Steidel} {et~al.}(2014){Steidel}, {Rudie}, {Strom}, {Pettini},  {Reddy}, {Shapley}, {Trainor}, {Erb}, {Turner}, {Konidaris}, {Kulas}, {Mace},  {Matthews}, \& {McLean}}]{Steidel14} {Steidel}, C.~C. {et~al.} 2014, \apj, 795, 165.

\bibitem[{{Veilleux} \& {Osterbrock}(1987)}]{VO87} {Veilleux}, S., \& {Osterbrock}, D.~E. 1987, \apjs, 63, 295

\bibitem[Vogt et al.(2014)]{Vogt14}Vogt, F.~P.~A.; Dopita, M.~A.; Kewley, L.~J.; Sutherland, R.~S.; Scharwaechter, J., Basurah, H.~M.; Ali, A., Amer, M.~A. 2014, \apj , 793, 127

\bibitem[{{Wuyts} {et~al.}(2014){Wuyts}, {Kurk}, {F{\"o}rster Schreiber},  {Genzel}, {Wisnioski}, {Bandara}, {Wuyts}, {Beifiori}, {Bender}, {Brammer}, {Burkert}, {Buschkamp}, {Carollo}, {Chan}, {Davies}, {Eisenhauer}, {Fossati}, {Kulkarni}, {Lang}, {Lilly}, {Lutz}, {Mancini}, {Mendel}, {Momcheva}, {Naab},  {Nelson}, {Renzini}, {Rosario}, {Saglia}, {Seitz}, {Sharples}, {Sternberg},  {Tacchella}, {Tacconi}, {van Dokkum}, \& {Wilman}}]{Wuyts14}{Wuyts}, E. {et~al.} 2014, \apjl, 789, L40

\bibitem[{{Yabe} {et~al.}(2012){Yabe}, {Ohta}, {Iwamuro}, {Yuma}, {Akiyama},  {Tamura}, {Kimura}, {Takato}, {Moritani}, {Sumiyoshi}, {Maihara},  {Silverman}, {Dalton}, {Lewis}, {Bonfield}, {Lee}, {Curtis Lake}, {Macaulay},  \& {Clarke}}]{Yabe12} {Yabe}, K. {et~al.} 2012, \pasj, 64, 60

\bibitem[{{Zahid} {et~al.}(2013{\natexlab{a}}){Zahid}, {Kashino}, {Silverman},  {Kewley}, {Daddi}, {Renzini}, {Rodighiero}, {Nagao}, {Arimoto}, {Sanders},  {Kartaltepe}, {Lilly}, {Maier}, {Capak}, {Carollo}, {Chu}, {Hasinger},  {Ilbert}, {Kajisawa}, {Koekemoer}, {Kovac}, {Le Fevre}, {Masters},  {McCracken}, {Onodera}, {Scoville}, {Strazzullo}, {Sugiyama}, {Taniguchi}, \&  {The COSMOS Team}}]{Zahid14}{Zahid}, H.~J. {et~al.} 2013{\natexlab{a}}, ArXiv e-prints, 1310.4950

\bibitem[{{Zahid} {et~al.}(2013{\natexlab{b}}){Zahid}, {Yates}, {Kewley}, \&  {Kudritzki}}]{Zahid13} {Zahid}, H.~J., {Yates}, R.~M., {Kewley}, L.~J., \& {Kudritzki}, R.~P.  2013{\natexlab{b}}, \apj, 763, 92, 1211.7062

\end{thebibliography}
\end{document}